\documentclass[doublecol]{epl2}

\usepackage{amsmath}
\usepackage{graphicx}
\usepackage{amsfonts}
\usepackage{amssymb}
\usepackage{epsfig}
\usepackage{color}
\usepackage{psfrag}
\usepackage{epstopdf}

\newcommand{\EQ}{\begin{equation}}
\newcommand{\EN}{\end{equation}}
\newcommand{\be}{\begin{equation}}
\newcommand{\ee}{\end{equation}}
\newcommand{\bea}{\begin{eqnarray}}
\newcommand{\eea}{\end{eqnarray}}

\title{Bulk and boundary effects on the decay of the thermodynamic Casimir force}

\author{Gesualdo Delfino\inst{1,2} \and Alessio Squarcini\inst{1,2}}

\institute{                    
  \inst{1}  SISSA - Via Bonomea 265, 34136 Trieste, Italy\\
  \inst{2}  INFN sezione di Trieste
}

\abstract{We consider the decay of the thermodynamic Casimir force in phases with a finite correlation length. For the case of the strip, we use properties of low energy two-dimensional field theory to show that the decay depends on the symmetry properties of the boundary conditions, in distinctive ways that we determine exactly. Features characteristic of the bulk universality class may induce modifications that we also discuss. Symmetry breaking and symmetry preserving boundary conditions exchange their role with respect to the decay of the force when exchanging spontaneously broken with disordered phases. Several of our arguments extend to higher dimensions.}

\begin{document}

\maketitle

\newpage
The quantum-electrodynamical Casimir force \cite{Casimir} is known to possess a thermodynamical analogue induced by the spatial confinement of the thermal fluctuations of a medium close to a second order transition point  \cite{FdG}. Such a thermodynamic (very often also called {\em critical}) Casimir force is observed experimentally \cite{GC,FYP,GSGC,HGDB,SZHHB,Bonn,NHB} and is important for a variety of applications to microdevices. Despite their relevance, on the other hand, theoretical characterizations have proved to be quite challenging, complicated as they are by the need to deal with interacting theories and by an essential dependence on boundary conditions. For the simplest geometry, a $D$-dimensional slab whose infinite boundary planes are separated by a distance $R$, and assigned uniform boundary conditions, it follows on general scaling grounds that the force (in temperature units $k_BT$ and per unit cross-sectional area) is $R^{-D}$ times a scaling function $\vartheta(R/\xi)$, where $\xi$ is the bulk correlation length\footnote{We refer to phases with finite correlation length.}. This function is universal, in the sense that it only depends on the symmetry of the order parameter, on $D$ and on the boundary conditions, but otherwise little is known in general about it, to the point that even the sign of the force represents a non-trivial problem. Indeed, while reflection positivity ensures that mirror symmetric boundaries with identical boundary conditions attract \cite{KK,Bachas}, the force is found to be repulsive in main instances of different conditions on the two boundaries (see e.g. \cite{FYP,VGMD} for experimental and numerical data, respectively, for the three-dimensional Ising universality class). On the other hand, it was pointed out in \cite{SD} that for different boundary conditions a tuning of boundary parameters can lead to the reversal of the force as $R/\xi$ varies, a circumstance neatly illustrated in \cite{AM} through exact computations for the Ising model in a strip.

In such an intricated situation, a general investigation of the function $\vartheta(R/\xi)$ can only start from asymptotics. For $R/\xi\to 0$ the force behaves as $\vartheta(0)/R^D$, and all the information about the boundary conditions is contained in the amplitude. Since this is a scale invariant limit for the bulk, boundary conformal field theory \cite{Cardy} allowed the exact determination of critical Casimir amplitudes in $D=2$ for several universality classes and scale invariant boundary conditions \cite{Cardy_strip,BX}. In this paper we consider the opposite limit $R/\xi\gg 1$. In the case $D=2$, that we study in detail, we show that the force decays differently for $R\gg\xi$ depending on the symmetry properties of the boundary conditions on the edges of the strip. Then, in principle, measuring the force in this limit provides a way to distinguish classes of boundary conditions realized in the physical system. Moreover, we show that the effect on the decay of symmetry breaking and symmetry preserving boundary conditions is interchanged when exchanging spontaneously broken with disordered phases. In recent years two-dimensional near critical behavior has been identified even in biological systems, such as cellular membranes \cite{bio1}, and the role of the thermodynamic Casimir force in this context has been investigated in \cite{bio2}. At the end of the paper we discuss to which extent our arguments extend to higher dimensions.

\vspace{.2cm}
We begin our analysis considering a two-dimensional statistical system confined on a strip of vertical width $R$ and length $L\to\infty$, with boundary conditions that we denote by $u$ on the upper edge and $d$ on the lower edge. The Casimir force per unit length between the two edges is given by
\EQ
{\cal F}_{ud}=\frac{1}{L}\partial_R\ln Z_{ud}=\frac{1}{L}\frac{\partial_RZ_{ud}}{Z_{ud}}\,,
\label{force}
\EN
where $-\ln Z_{ud}$ is the contribution to the free energy due to the interaction between the edges. The system is close to a second order phase transition point, so that its scaling limit corresponds to a Euclidean field theory, which in turn can be regarded as the analytic continuation to imaginary time of a relativistic quantum field theory in one spatial dimension. If $H$ denotes the Hamiltonian of this quantum theory, the partition function $Z_{ud}$ can be written as
\EQ
Z_{ud}=\langle B_u|e^{-HR}|B_d\rangle\,,
\label{pf}
\EN
where $|B_d\rangle$ and $|B_u\rangle$ are boundary states specifying the initial and final conditions of the imaginary time evolution{; they can be expanded over the complete basis of asymptotic particle states of the bulk ($R=\infty$) theory, which are eigenstates of the Hamiltonian $H$. 

We consider uniform, i.e. translation invariant, boundary conditions. The use of translation invariant boundary states\footnote{{See \cite{DV2,DS} for the non-translation-invariant case.}} in the off-critical case was illustrated in \cite{GZ} and exploited for free energy calculations on the strip in \cite{LcMSS} in the context of integrable field theories. A study of the leading finite size effects was then performed in \cite{BPT1,BPT2}, with particular attention for the precise relation between boundary state amplitudes and scattering amplitudes in the ``crossed channel''. In the present paper we are interested in the way the symmetry properties of the boundary conditions affect the finite size dependence in the different phases of the system and for the different universality classes, a subject whose systematic study was initiated in \cite{DV1} in the context of crossing probabilities in percolation; but for Eq.~(\ref{f_abc}), that we borrow from \cite{GZ}, our derivations are self-contained.}  

The nature of bulk excitations differs above and below the critical temperature $T_c$ associated to the spontaneous breaking of the symmetry (corresponding to a group $G$) characterizing the universality class; we discuss first the case $T<T_c$. Then, in two dimensions, the system possesses discrete degenerate ground states, corresponding to degenerate vacua of the associated quantum theory, that we denote by $|\Omega_a\rangle$, $a=1,\ldots,n$. For topological reasons, the elementary excitations are kinks $|K_{ab}(\theta)\rangle$ interpolating between different vacua $|\Omega_a\rangle$ and $|\Omega_b\rangle$; the rapidity $\theta$ parameterizes the energy and momentum of these relativistic particles as $(e,p)=(m\cosh\theta,m\sinh\theta)$, where $m$ is the kink mass. In general the kink mass depends on the indices $a$ and $b$; here, however, we will be interestend only in the leading large distance behavior of the Casimir force, which is determined by the particles with the lowest mass, and for this reason we will keep track only of the lightest kinks. Similarly, among the bound states that kinks may form, we will be interested in those arising in the topologically neutral channels $|K_{ab}(\theta_1)K_{ba}(\theta_2)\rangle$, and will denote by $|B_{a}(\theta)\rangle$ the lightest among them, with mass $m_B<2m$. Throughout the paper we call ``exponential'' correlation length and denote by $\xi$ the correlation length defined by the large distance decay $r^{-\alpha}e^{-r/\xi}$ of the order parameter two-point function in the bulk theory. Since the order parameter operator is topologically neutral, $\xi$ is $1/2m$ in absence of neutral bound states, and $1/m_B$ otherwise. 

The boundary conditions on the edges of the strip can be either symmetry preserving (i.e. left invariant by the action of the group $G$) or symmetry breaking. In the latter case we consider symmetry breaking (by a boundary field $h$) in favor of one of the degenerate vacua $|\Omega_a\rangle$, and denote by $|{\cal B}_a(h)\rangle$ the corresponding boundary state. The expansion over bulk states of one such boundary state, say $|{\cal B}_1(h)\rangle$, will be of the form
\bea
|{\cal B}_1(h)\rangle &=& |\Omega_1\rangle+g(h)|B_{1}(0)\rangle 
\label{B1}\\
&+& \sum_{b\neq 1}\int\frac{d\theta}{2\pi}f_{b}(\theta,h)|K_{1b}(-\theta)K_{b1}(\theta)\rangle+\cdots\,,\nonumber
\eea
where the bulk states start and end on the vacuum $|\Omega_1\rangle$, and have zero total momentum as a consequence of translation invariance of the boundary condition; the dots stay for states with higher total mass\footnote{To be definite, we discuss the case $m_B>m$.} whose contribution to the large distance expansion of the Casimir force is subleading. Turning to the symmetry preserving boundary states, we will denote them by $|{\cal B}_0(u)\rangle$, with $u$ collectively denoting boundary parameters. These states expand in the form
\bea
|{\cal B}_0(u)\rangle &=& \sum_{a}\left\{v_a(u)|\Omega_a\rangle+g_{a}(u)|B_{a}(0)\rangle\right. 
\label{B0}\\
&+& \left.\sum_{b\neq a}\int\frac{d\theta}{2\pi}f_{aba}(\theta,u)|K_{ab}(-\theta)K_{ba}(\theta)\rangle \right. \nonumber\\
&+& \sum_{c\neq a}\left[g_{ac}(u)|K_{ac}(0)\rangle\right.\nonumber\\
&+& \sum_{b\neq a,c}\int\frac{d\theta}{2\pi}f_{abc}(\theta,u)|K_{ab}(-\theta)K_{bc}(\theta)\rangle]\}+.., \nonumber
\eea
with the different vacua treated on the same footing.

We can now consider the large $R$ asymptotics of the Casimir force for the different combinations of boundary conditions (\ref{B1}) and (\ref{B0}). For symmetry preserving, or {\em free}, boundary conditions on both edges the leading contribution comes from the single-kink state in (\ref{B0}), and we have
\bea
Z_{00} &=& \langle{\cal B}_0(u)|e^{-HR}|{\cal B}_0(u')\rangle \label{Z00}\\
&\sim & \sum_a\left[v_a^*(u)v_a(u')+mL\sum_{c\neq a}g_{ac}^*(u)g_{ac}(u') e^{-mR}\right],\nonumber
\eea
where we used $\langle\Omega_a|\Omega_b\rangle=\delta_{ab}$, $\langle K_{ab}(\theta)|K_{ac}(\theta')\rangle=2\pi\delta(\theta-\theta')\delta_{bc}$ and $2\pi\delta(0)=mL$. Equation (\ref{force}) then gives
\EQ
{\cal F}_{00}\sim -A_{00}\,m^2\,e^{-mR}\,,
\label{F00}
\EN
with $A_{00}=\sum_{a,c\neq a}g_{ac}^*(u)g_{ac}(u')/\sum_av_a^*(u)v_a(u')$. 

For boundary conditions ${\cal B}_1(h)$ on the upper edge and ${\cal B}_1(h')$ on the lower edge, the two-kink state gives the leading contribution to the force in absence of neutral bound states ($g=0$ in (\ref{B1})). The eigenvalue of $e^{-HR}$ on the two-kink state is $e^{-2mR\cosh\theta}$, so that the limit of large $mR$ is determined by the behavior of the excitations at small rapidities, which is a property of the bulk theory. With few exceptions, interacting particles in 1+1 dimensions behave at low energies as free fermions, and here we will discuss this generic case. Then for the product of states entering (\ref{pf}) we have in this limit
\bea
&&\hspace{-1cm} \langle K_{1c}(\theta')K_{c1}(-\theta')|K_{1b}(-\theta)K_{b1}(\theta)\rangle\\
&\sim & \delta_{bc}(2\pi)^2\{[\delta(\theta-\theta')]^2-[\delta(\theta+\theta')]^2\}\nonumber\\
&=& \delta_{bc}\,2\pi mL\cosh\theta\,[\delta(\theta-\theta')-\delta(\theta+\theta')]\,.\nonumber
\eea
A further consequence of the low energy fermionic statistics is that the two-kink amplitudes in (\ref{B1}) vanish at $\theta=0$ (namely when the two particles have the same momentum), and can be written at small rapidity as
\EQ
f_b(\theta,h)\sim C_b(h)\,\theta\,.
\label{f_b}
\EN 
The last two equations allow us to calculate the two-kink contribution to the partition function $Z_{11}=\langle B_1(h)|e^{-HR}|B_1(h')\rangle$ in the large $R$ limit as 
\EQ
A_{11}2mL\int\frac{d\theta}{2\pi}\,\theta^2 e^{-2mR(1+\theta^2/2)}=\frac{A_{11}}{2\sqrt{\pi}}\frac{mL}{(mR)^{3/2}}\,e^{-2mR}\,,
\EN
with $A_{11}=\sum_{b\neq 1}C_b^*(h)C_b(h')$. The corresponding force is then
\EQ
{\cal F}_{11}\sim-\frac{A_{11}}{\sqrt{\pi}}\frac{m^2}{(mR)^{3/2}}\,e^{-2mR}\,,
\label{F11}
\EN
in absence of topologically neutral bound states, and $-g^*(h)g(h')m_B^2\,e^{-m_BR}$ if such a bound state is present\footnote{{The single-particle contribution to the free energy has been investigated in \cite{BPT1,BPT3}, where its amplitude, including the sign, has been determined for some integrable field theories.}}. Notice that the force is attractive for $h=h'$, as for ${\cal F}_{00}$ with $u=u'$; this agrees with the general result for identical mirror symmetric boundary conditions. Apart from these two cases, the sign is not determined in general.

If we consider boundary conditions ${\cal B}_1(h)$ on the upper edge and ${\cal B}_0(u)$ on the lower edge, the calculation proceeds as in the previous case, with one important difference. It was found in \cite{GZ} that when a boundary state contains a two-particle contribution such that the two particles individually contribute single-particle states to the expansion, then the amplitude of the two-particle state has a simple pole at $\theta=0$. For the state (\ref{B0}) this means in particular that for small rapidity
\EQ
f_{abc}(\theta,u)\sim\frac{C_{abc}(u)}{\theta}\,,
\label{f_abc}
\EN
with $C_{abc}\propto g_{ab}g_{bc}$; this is still consistent with low energy fermionic statistics since $f_{aba}(\theta,u)$ changes sign when the momenta of the two particles are interchanged ($\theta\to-\theta$). It follows from the combination of (\ref{f_b}) and (\ref{f_abc}) that the two-kink contribution to the partition function $Z_{10}$ for $mR$ large reads
\EQ
A_{10}\,2mL\int\frac{d\theta}{2\pi}\,e^{-2mR(1+\theta^2/2)}=\frac{A_{10}\,mL}{\sqrt{\pi\,mR}}\,e^{-2mR}\,,
\EN
with $A_{10}=\sum_{b\neq 1}C_{b}^*(h)C_{1b1}(u)$. The force is then
\EQ
{\cal F}_{10}\sim-\frac{2A_{10}\,m^2}{\sqrt{\pi\,mR}}\,e^{-2mR}\,,
\label{F10}
\EN
in absence of neutral bound states, and $-g^*(h)g_1(u)m_B^2\,e^{-m_BR}$ otherwise; in writing ${\cal F}_{10}$ we are choosing the normalization with $v_1=1$ for the boundary state (\ref{B0}).

A last possible choice of uniform boundary condition is to take  ${\cal B}_1(h)$ on the upper edge and ${\cal B}_2(h')$ on the lower edge, with the latter choice corresponding to symmetry breaking in the direction of a different vacuum $|\Omega_2\rangle$. It follows from (\ref{B1}) that in this case the two boundary states have zero overlap, so that the free energy $-\ln Z_{12}$ is infinite. This corresponds to the fact that, in our large $R$ limit, the boundary conditions we are considering lead to phase separation, with an interfacial tension equal to the kink mass $m$ \cite{DV2} and an excess free energy $mL$ which diverges as $L\to\infty$. 

We can now consider the case $T>T_c$ of unbroken bulk symmetry. In this case the bulk theory possesses a single, symmetry invariant vacuum $|\Omega\rangle$, and the elementary excitations are no longer topological. In general, they will form a multiplet of particles $A_i$, with mass $\tilde{m}$, transforming according to a representation of the symmetry group $G$. These particles may give rise to bound states, and we denote by $B$ the lightest among those invariant under the action of the group.
Normally in a disordered phase the components of the order parameter operator create the elementary excitations $A_i$, so that the exponential correlation length is $\xi=1/\tilde{m}$. Concerning the expansion of boundary states on bulk states, it is natural to consider neutral and charged boundary states. Neutral boundary states are those unaffected by the action of the group, and expand as
\bea
|\tilde{\cal B}_0(u)\rangle &=& |\Omega\rangle+\gamma(u)|B(0)\rangle
\label{B0high}\\
&+&\sum_{ij}\int\frac{d\theta}{2\pi}f_{ij}(\theta,u)|A_{i}(-\theta)A_{j}(\theta)\rangle+\cdots\,,\nonumber
\eea
where the tilde is used to distinguish (\ref{B0high}) from the expansion (\ref{B0}) below $T_c$. Comparison with (\ref{B1}) then shows that the derivation of the large $R$ behavior of ${\cal F}_{00}$ above $T_c$ retraces that of ${\cal F}_{11}$ below $T_c$. The charged state $|{\cal B}_i\rangle$, depending on some boundary parameter $\lambda$, transforms as the particle $A_i$ under the action of the group, and expands as
\EQ
|{\cal B}_i(\lambda)\rangle=|A_i(0)\rangle+\sum_{jk}\int\frac{d\theta}{2\pi}f_{ijk}(\theta,\lambda)|A_{j}(-\theta)A_{k}(\theta)\rangle+..\,.
\label{Bi}
\EN
Notice that $Z_{ii}=\langle{\cal B}_i|e^{-HR}|{\cal B}_i\rangle\propto \tilde{m}Le^{-\tilde{m}R}$ for $R$ large, so that ${\cal F}_{ii}\sim-\tilde{m}/L$. We see that the absence of the vacuum contribution in (\ref{Bi}) makes $L{\cal F}_{ii}$ non-extensive in $L$ and non-vanishing as $R\to\infty$. Since extensivity and large $R$ suppression should be preserved by the boundary state $|\tilde{\cal B}_1(h)\rangle$ corresponding to the presence of a symmetry breaking boundary field (the analogue of (\ref{B1}) for $T>T_c$), we are led to conclude that this is realized by a superposition of (\ref{B0high}) and (\ref{Bi}). Comparing such a superposition to (\ref{B0}) we see that the derivation of the large $R$ behavior of the Casimir force above $T_c$ for symmetry breaking boundary fields acting on both edges retraces that of ${\cal F}_{00}$ below $T_c$.

The dynamics of bulk excitations is known exactly for most universalilty classes in two dimensions. For example, in the $q$-state Potts model \cite{Wu}, which exhibits a second order transition for $q$ up to 4, the high and low temperature phases are related by duality, and have the same mass spectrum, with the same mass $m$ for the kinks below $T_c$ and the particles above. These are the only excitations for $q=2,3$, while a neutral bound state with mass $\sqrt{3}m$ exists for $q=4$ \cite{CZ} and affects the Casimir force in the way we described. The case $q=3$ provides one of the exceptions we mentioned to the fermionic low energy behavior of bulk excitations, and this results in modifications of (\ref{F11}) and (\ref{F10}) that we will detail elsewhere. For the Ising model ($q=2$), the exact relations satisfied by the Casimir force in the strip when exchanging high with low temperature and, simultaneously, fixed with free boundary conditions \cite{AM,ES}, are a duality-enhanced example of the correspondences we obtained above. Similarly, it follows from our analysis that, for the Ising model with free boundary conditions on both edges of the strip, the Casimir force has the asymptotic form (\ref{F00}) below $T_c$ and (\ref{F11}) above; this accounts for the asymmetry of the force across $T_c$ studied on the lattice in \cite{AM2}.

Taking as an additional example the XY universality class, characterized by $O(2)$ symmetry, we need to remember that continuous symmetries cannot break spontaneously in two dimensions \cite{MW,H}, and that the transition is of the Berezinsky-Kosterlitz-Thouless type \cite{BKT}. While the low temperature phase renormalizes onto a conformal field theory, our results for massive phases apply above the transition temperature. This disordered phase is described by a field theory with fermionic low energy behavior and without bound states \cite{Zamo_O(n)}. Hence, we find in particular the asymptotic result ${\cal F}_{00}\propto R^{-3/2}e^{-2mR}$, which can be compared to $R^{(1-D)/2}\exp(-2R/\hat{\xi})$ obtained in \cite{KD} from a pertubative calculation in $D=4-\epsilon$ dimensions. It is not surprising that the $\epsilon$-expansion does not reproduce for $D=2$ the prefactor $R^{-3/2}$, which originates from the non-perturbative property (\ref{f_b}). Concerning the exponential factor, $\hat{\xi}$ should be identified with $\xi=1/m$.

Several of the arguments used for the strip can be generalized to the case $D>2$. Boundary states now describe boundary conditions on $(D-1)$-dimensional hyperplanes, and can still be expanded on the asymptotic states of the bulk theory {\cite{BPT2,BPT3,vortex}}. In $D>2$ also continuous symmetries can break spontaneously, but the presence of massless (Goldstone) particles in the expansion of the boundary states will prevent exponential decay of the force below\footnote{See the profile of the force determined in \cite{DGHHRS} for the three-dimensional $O(n\to\infty)$ case.} $T_c$. For phases with spontaneously broken discrete symmetry the force still decays exponentially, but the elementary particle excitations are no longer kinks and the analysis differs substantially from the case $D=2$. The symmetry considerations we made above for the case $T>T_c$ should instead hold in general. A limitation for the asymptotic analysis in $D>2$ is that the low energy behavior of the amplitudes of two-particle states (the analogue of (\ref{f_b}) and (\ref{f_abc}) above) is not known; in principle simulation results for the force can be used to investigate this point. On the other hand, when the decay of the force ${\cal F}_{ud}=L^{1-D}\partial_R\ln Z_{ud}$ is ruled by a single-particle term, the large $R$ suppression is $D$-independent. For example, for the $O(n)$ model with a boundary field $h$ on both boundaries, the force is expected to decay as $\alpha_D(h)\,\xi^{-D}e^{-R/\xi}$ above $T_c$, with $\alpha_D(h)$ a pure number and $\xi$ the exponential correlation length.

\vspace{.2cm}
In summary, we studied the decay of the thermodynamic Casimir force on an infinitely long strip whose width $R$ is much larger than the bulk correlation length. The analysis exploits the expression of the boundary conditions in terms of the particle excitations of the bulk theory. Using low energy properties of two-dimensional field theory we determined the exact form of the large $R$ suppression, and showed that it depends in distinctive ways on the symmetry properties of the boundary conditions. The possibility to detect symmetry classes of boundary conditions from the functional form of the decay of the force contrasts with what happens in the opposite limit ($R$ much smaller than the correlation length), in which boundary conditions only affect numerical amplitudes. We also discussed which features specific of the bulk universality class may affect the decay of the force. The different nature of the bulk excitations above and below the critical temperature was shown to induce in general a different behavior of the force in the two regimes. On the other hand, the large $R$ suppression does not change when exchanging spontaneously broken with disordered phases and, at the same time, symmetry breaking with symmetry preserving boundary conditions, a circumstance that must be regarded as a weaker, but more general, version of duality relations known for the Ising model. The formalism makes transparent that the sign of the force at large $R$ depends on the boundary parameters if these are different on the two edges of the strip, and is attractive if they are identical. Finally we discussed how several of our arguments extend to higher dimensions and yield specific predictions.

\end{document}